\documentstyle[12pt,oldlfont]{article}
\def\beq{\begin{equation}}
\def\eeq{\end{equation}}
\def\bea{\begin{eqnarray}}
\def\eea{\end{eqnarray}}
\def\bq{\begin{quote}}
\def\eq{\end{quote}}

\def\NP{{\it Nucl.Phys.} }
\def\PL{{\it Phys.Lett.} }
\def\PR{{\it Phys.Rev.} }

\def\SJPN{{\it Soviet J.Part.Nucl.} }

\def\TMP{{\it Theor.Math.Phys.} }

\def\gappeq{\mathrel{\rlap {\raise.5ex\hbox{$>$}}
{\lower.5ex\hbox{$\sim$}}}}

\def\lappeq{\mathrel{\rlap{\raise.5ex\hbox{$<$}}
{\lower.5ex\hbox{$\sim$}}}}

\begin{document}
\topmargin -0.5cm
\oddsidemargin -0.3cm
\pagestyle{empty}
\begin{flushright}
{CERN-TH.7226/94}
\end{flushright}
\vspace*{5mm}
\begin{center}
{\bf SUPERGRAVITY BEFORE AND AFTER 1976} \\
\vspace*{1cm}
{\bf D.V. Volkov}$^{*)}$ \\
\vspace{0.3cm}
K.F.T.I., Kharkov, Ukraine \\
and \\
Theoretical Physics Division, CERN \\
CH - 1211 Geneva 23 \\
\vspace*{2cm}
{\bf ABSTRACT} \\ \end{center}
\vspace*{5mm}
\noindent
This paper is  part of the lecture given at the TH Division of CERN and
devoted to the CXXV anniversary of the birthday of Elie Cartan
(1869-1951). It is shown how the methods of differential geometry, due to
E. Cartan, were applied to the construction of the supersymmetry
transformation law and to the actions for Goldstone fermions and
supergravity.

\vspace*{5cm}
\noindent
\rule[.1in]{16.5cm}{.002in}

\noindent
$^{*)}$ Permanent address: Kharkov Institute of Physics and Technology,
Kharkov 310108,  Ukraine. e-mail address: dvolkov @ kfti.kharkov.ua.
\vspace*{0.5cm}

\begin{flushleft} CERN-TH.7226/94 \\
April 1994
\end{flushleft}
\vfill\eject

\setcounter{page}{1}
\pagestyle{plain}

In the review articles and monographs on supergravity,  its
discovery is usually dated as 1976. However, a number of papers on
supergravity appeared before 1976, beginning from 1972 \cite{aaa}-\cite{ee}.
As has now been confirmed, those papers are directly related to the
recognized version of supergravity \cite{ff},\cite{ggg} and in some sense
were the starting points for the later ones. In this paper, I will try to
fill in the historical gap (1972-1976).

Supergravity is the gauged version of the global supersymmetry. Therefore I
will begin by exposing briefly those elements of  supersymmetry which
will be essential for the following.

Supersymmetry has been independently discovered by three groups of authors:
Yu. Gol'fand and E. Lichtman \cite{hh}; D. Volkov and V. Akulov \cite{aaa};
J. Wess and B. Zumino \cite{jj}. The motivations and starting points  used
by the three groups of authors were quite different.

In Ref. \cite{hh} the motivation was to introduce a parity violation
into the quantum field theory. The starting point of the papers
\cite{aaa},\cite{bb} was the question whether Goldstone particles with
spin one-half might exist. The authors of Ref. \cite{jj}  made the
generalization of the supergroup which first appeared in the
Neveu-Ramond-Schwarz dual model \cite{kk},\cite{lll} to the
four-dimensional world.

The approach of the papers \cite{aaa},\cite{bb} was the most appropriate
for gauging the super-Poincar\'e group which was done a little later in the
papers of D. Volkov and V. Soroka \cite{cc},\cite{dd} (1973-1974), where the
super-Higgs effect in  supergravity was elaborated.

The connection of the papers \cite{aaa},\cite{bb} and \cite{cc},\cite{dd}
is very natural, as in the gauge field theories the transformation law for
the gauge fields is determined by the same group structure which gives a
description of the Goldstone fields.
So I will consider, as an introduction, those features of the supersymmetry
theory that are essential for its gauging.

A detailed exposition of the route along which the generalization of the
Poincar\'e group to the super-Poincar\'e group was made is contained in
Ref. \cite{bb}. As this paper is not well known, I will briefly recall its
most essential points.

As has been mentioned above, the starting point was the question whether
Goldstone particles with spin one-half might exist. At the end of the
Sixties, the method of the phenomenological Lagrangians for the description
of Goldstone particles, so that it reproduced the results of PCAC
(partially conserved axial-vector currents) and the current algebra, had
been invented by S. Weinberg and J. Schwinger. At the time of the XIVth
Conference on high-energy physics (Vienna, 1968), the problem of the
current algebra and of the phenomenological Lagrangians had been
intensively discussed (see Weinberg's rapporteur talk \cite{mm}). There were
two papers presented in the current algebra section of the conference in
which the generalization of the method of phenomenological Lagrangians to
an arbitrary internal symmetry group had been elaborated. One paper was
presented by B. Zumino \cite{nn} (co-authors C. Callan, S. Coleman and J.
Wess), and another one by myself \cite{oo}. The main results of the papers
were  practically identical. The difference was that in Ref. \cite{oo}
\footnote{In \cite{pp}, which contains \cite{oo}, the methods of E. Cartan
was also used for the construction of phenomenological Lagrangians for the
spontaneously broken symmetry groups, containing the Poincar\'e group as a
subgroup. The Lagrangian for the Goldstone fermions $(9)$ is an example of
such a construction.} the works of E. Cartan on symmetric spaces and his
method of the exterior differential forms was intensively used.

In both papers, as well as in E. Cartan's works \footnote{And also in the
many papers that followed.}, the decomposition of a group $G$ on the factors
\beq
G = KH
\label{1}
\eeq
was used, with the parameters of the coset $K$ forming a homogeneous
space, and $H$ being the holonomy group of the space.

In the method of the phenomenological Lagrangians, the co-ordinates of the
coset $K$ correspond to the Goldstone fields.  Therefore the quantum
numbers of Goldstone fields coincide with the quantum numbers of the
generators of the coset $K$. This fact gives the answer to the question
about the possibility of the existence of Goldstone particles with spin
one-half.

To ensure such a possibility, the Poincar\'e group should be generalized in
such a way that the generalization contains the generators with spin
one-half and with commutation relations corresponding to the Fermi
statistics. From a technical point of view the problem was what
representation of the Poincar\'e group is the most appropriate for such a
generalization.

The solution of this technical problem was that the following
representation of the Poincar\'e group
\beq
G_{\rm Poincare} =
\left(\matrix{L &iXL^{+ -1}\cr 0 & L^{+ -1}}\right) =
\left(\matrix{1 & iX \cr 0 & L}\right)~~
\left(\matrix{L & 0\cr 0&L^{+ -1}}\right)~,
\label{2}
\eeq
where $L, L^{+ -1}$ and $X$ are the
2$\times$2 matrices $L = L_{\alpha}^{\phantom{\alpha}\beta}; L^{+ -1} =
L^{\dot\alpha}_{\phantom{\dot\alpha}\dot\beta}; X = X_{\alpha\dot\beta}$,
had all required properties.

In the generalization to the super-Poincar\'e group $K_{\rm transl.}$ plays
the main role.

Let us write it as consisting  of four blocks:
\beq
K = \left(
\begin{tabular}{c|c}
1 & iX \\ \hline
 0 & 1 \\
\end{tabular}
\right)
\label{3}
\eeq

Separating the blocks as follows
\beq
K^{\prime} = \left(
\begin{tabular}{c|c|c}
1 & -- & iX \\
\hline
0 & 1 & -- \\
\hline
0 & 0 & 1 \\
\end{tabular}
\right)
\label{4}
\eeq
one can insert into the newly formed hatched blocks  Grassmann spinors
$\theta_{\alpha}$ and $\bar\theta_{\dot\alpha}$ so that
$K^{\prime}$ becomes
\beq
K^{\prime} = \left(
\begin{tabular}{c|c|c}
1 & $\theta$ & iX$^{\prime}$ \\
\hline
0 & 1 & $\bar\theta$ \\
\hline
0 & 0 & 1 \\
\end{tabular}
\right)
\label{5}
\eeq

The matrix $K^{\prime}$ forms a group, but only under the condition that
$X^{\prime}$ is complex. To satisfy the reality condition for $X^{\prime}$
with the reality condition for $X$ in (\ref{2}) and simultaneously
conserving the group properties of (\ref{5}), the following representation
of $X^{\prime}$, as a sum of real and imaginary parts, is appropriate
$$
iX^{\prime} = iX + {1\over 2} \theta\bar\theta~.
$$

The resulting expression for the super-Poincar\'e group is the following
\beq
G_{\rm SUSY} =
\left(\matrix{
1 & \theta & iX+{1\over 2}\theta\bar\theta \cr
0 & 1 & \bar\theta \cr
0 & 0 & 1}\right)~~
\left(\matrix{
L & 0 & 0 \cr
0 & 1 & 0 \cr
0 & 0 & L^{+-1}}\right)
\label{6}
\eeq

{}From (15) one gets the transformation law for the superspace coordinates
\bea
X^{\prime} &=& X + i\epsilon\bar\theta \nonumber \\
\theta^{\prime} &=& \theta + \epsilon~;~~\bar\theta^{\prime} = \bar\theta +
\bar\epsilon
\label{7}
\eea
as well as the following expressions for the left-invariant vielbein
one-differential forms on the superspace $(X,\theta)$
\bea
e^a &=& dX^a + i\bar\theta \gamma^a d0\nonumber \\
e^{\alpha} &=& d\theta
\label{8}
\eea
The latter are received as components of $K^{-1}dK$ corresponding to the
generators of $K$ -- now being the supertranslation subgroup of (\ref{6}).

The action for the Goldstone fermions is an integrated four-form pulled back
onto the four-dimensional Minkowski space (the world space) and
\beq
L_{GF} = {1\over 24} ~~\epsilon_{abcd} e^a e^b e^c e^d
\label{9}
\eeq

The expressions (\ref{7})-(\ref{9}) were written in \cite{aaa}
without the detailed deduction which was later reproduced in \cite{bb}.

Now, as a first step in discussing the problem of supersymmetry gauging, I
cite the final sentence of \cite{aaa}:
\begin{quote}
``...the gravitational interaction may be included by means of introducing
the gauge fields for the Poincar\'e group. Note that if the gauge field for
the transformation (\ref{3}) [formula (7) of the present text]  is also
introduced, then as a result of the Higgs effect the massive gauge field
with spin three-halves appears and the considered Goldstone particle with
spin one-half disappears."
\end{quote}

Now let us go to the procedure of gauging the supersymmetry.

The gauge fields for the local supersymmetry group can be introduced in the
standard way
\beq
A_{\rm SUSY} (d) = \left(
\matrix{
\omega_{\alpha}^{\phantom{\alpha}\beta}(d) & \psi_{\alpha}(d) &
e_{\alpha\dot\beta}(d) \cr
& 1 & \bar\psi_{\dot\beta}(d) \cr
& & \omega^{\dot\alpha}_{\phantom{\dot\alpha}\dot\beta}(d)}
\right)
\label{10}
\eeq
with the standard transformation law
\beq
A^{\prime}(d) = G^{-1}A(d) G + G^{-1}dG
\label{11}
\eeq
where $A(d)$ is the $g_{\rm SUSY}$-algebra valued one-differential forms.

Expression (\ref{11}) may be
interpreted in two ways: the first one as the transformation law for $A(d)$,
and the second one in which $G$ or some of its coset $K$ is interpreted as
the space of Goldstone fields.

In the latter case, if the Goldstone fields transform as $G^{\prime}_L =
LG$ ($L$ is the left multiplication on the group $G$) then Eq. (\ref{11})
is invariant.

Analogously, if the coset $K$ contains the Goldstone fields, then
$K^{\prime}_L = LKH^{-1}(K,L)$
\beq
A^{\prime}(d) = H (K^{-1}AK + K^{-1}dK) H^{-1} + HdH^{\prime}
\label{12}
\eeq
so that the projections of $K^{-1}AK + K^{-1}dK$ on the generators of $K$
are the covariants of the subgroup $H$.

The second way of interpreting (\ref{11}) is appropriate for considering
the spontaneously broken supergravity; the first one is useful if pure (not
broken) supergravity is considered.

In our case, the forms
$e_{\alpha\dot\beta}(d)~\psi_{\alpha}(d)$ and $\bar\psi_{\dot\alpha}(d)$,
as well as the curvature tensor
$R^{\alpha}_{\phantom{\alpha}\beta}(d,d^{\prime})$;
$R^{\dot\alpha}_{\phantom{\dot\alpha}\dot\beta}(d,d^{\prime}_z)$ for the
Lorentz connection $\omega (d)$, are the covariants of the Lorentz subgroup
$L$.

The covariant forms (\ref{10}) in the presence of the Goldstone fields
$X^a, \theta^{\alpha}$ and $\theta^{\dot\alpha}$ may be written as
$$
\tilde e (d) = e(d) + {\cal D}X + i\bigg[(2\psi(d) + {\cal D}\theta )\bar\theta
\theta(2\bar\psi (d) + {\cal D}\bar\theta)\bigg]
\eqno{(13a)}
$$
$$
\tilde\psi (d) = \psi(d) + {\cal D}\theta
\phantom{xxxxxxxxxxxxxxxxxxxxxxxxxxxxx}
\eqno{(13b)}
$$
$$
\tilde\omega (d) = \omega (d)
\phantom{xxxxxxxxxxxxxxxxxxxxxxxxxxxxxxxxxx}
\eqno{(13c)}
$$
$$
\tilde R(d,d^{\prime}) = R(d,d^{\prime}) = d\omega (d^{\prime}) -
d^{\prime}\omega
\bigg[\omega (d), \omega (d^{\prime})\bigg]
\phantom{xxxxx}
\eqno{(13d)}
$$
Note that in the case of infinitesimal transformation only terms linear in
$\theta$  are present in (13).

One can now construct the following invariant four-differential forms
contracting the indices of the one-differential forms \cite{cc},\cite{dd}
$$
W_1  = R(d_1,d_2) e(d_3)~e(d_4)
 \eqno{(14a)}
$$
$$
W_2 = {\cal D}\bar\psi (d_1,d_2) e(d_3) \psi (d_4)
\eqno{(14b)}
$$
$$
W_3 = e(d_1) e(d_2) e (d_3) e (d_4)
\eqno{(14c)}
$$
$$
W_4 = \bar\psi (d_1) e(d_2) e(d_3) \psi (d_4)
\eqno{(14d)}
$$
\addtocounter{equation}{2}
using either the spinor notation $(\alpha ,\dot\alpha$- indices) or the
more usual vector notation for $R(d_1,d_2)$ and $e(d)$, so that
$W_1,W_2,W_3$ and $W_4$ represent correspondingly the Einstein action, the
Rarita-Schwinger kinetic term, the cosmological term and the mass term for
the Rarita-Schwinger field.

The resulting action is the sum
\beq
W = \underbrace{a_1W_1~~+~~a_2W_2}_{\matrix {\rm the~pure~SUGRA}} ~~\quad+
\underbrace{a_3W_3~~+~~a_4W_4}_{\matrix {\rm the~terms~due~to~the \cr \rm
sponta
breakdown~of \cr \rm the~super-Poincare~group}}
\label{15}
\eeq
The fact that the sum $a_1W_1 + a_2W_2$ is the pure unbroken supergravity
follows from counting the degrees of freedom of the Rarita-Schwinger field
with the action $W_2$ in a gravitational background. It is easy to show that
if the gravitational background satisfies the equations of motion for the
Einstein action $W_1$, then the Rarita-Schwinger field has two degrees of
freedom. So the Goldstone fields do not contribute on the mass shell of the
gravitational field.

The formulas (14) and (\ref{15}) and transformation law (13)
are the main results of the papers \cite{cc},\cite{dd}.

Now let us turn to the works on supergravity that appeared in 1976. The
first of these works \cite{ff} used
the second-order formalism for the Einstein action with
a rather complicated transformation law for the supergravity gauge fields. A
more simplified form of the action and transformation law has  been proposed
in the paper of Deser and Zumino \cite{ggg}. These
authors have written in their introduction:
\begin{quote}
``The key to our results lies in the use of the first-order formalism for
gravitation,
in which vierbeins and connection coefficients are treated independently.
Minimal coupling in this sense implies the existence of torsion, or of
non-minimal contact interactions in second-order language. The first-order
formulation with torsion is closely related to the description of
supergravity in superspace \cite{ee}
\footnote{The reference note is given
according to the list of references of the present paper. In the
paper referred to, E. Cartan's methods of differential geometry are firstly
generalized to the graded superspaces. It is also shown that ``the flat
superspace" has torsion, and that the holonomy group for the curvature in
the superspace formulation of supergravity should be the Lorentz group.}."
\end{quote}

The transformation laws for the gauge fields $e(d)$ and $\psi (d)$ which
were used in the paper of Deser and Zumino \cite{ggg} coincided with
(22a,b) but were different from (22c) for the Lorentz connection form
$\omega (d)$. The further development of the supergravity theory has shown
that the explicit forms of the $\delta\omega$ variation does not matter.

As a result of further investigations, the first-order formalism and the
gauged supergroup approach to the supergravity based on the
transformations (13) is now accepted to be the simplest way to
supergravity\footnote{See, for example, \cite{qq}.}. The important steps
along this line of reasoning were the rheonomy theory of supergravity as
well as the above-mentioned explicit proofs of the invariance of the
supergravity action by using the transformation law (13), and
showing that the condition $\delta\omega (d) = 0$ (13c)  considerably
simplifies the proof. The advantages of the first-order formalism were
proved to be useful in many aspects of the theory.

The period of intensive development of  supergravity since 1976 is well
described in the review article \cite{qq} and others which followed. In
the text of the review \cite{qq} there is no reference to the papers
\cite{aaa}-\cite{ee}. So the present paper may be considered as an addendum
to \cite{qq} with reference to the papers which were published before 1976
and were essentially based on the first-order formalism, developed, in its
main features, by E. Cartan.

\vspace*{2cm}
\noindent
{\bf Acknowledgements}

I am grateful to the Theoretical Physics Division for kind hospitality
during my stay at CERN. This work was partially supported by the
International Science Foundation and Ukrainian State Committee in Science
and Technologies, Grant Nr 2/100.

\vfill\eject

\end{document}